\documentclass[12pt]{iopart}
\usepackage{iopams}
\begin{document}

\title{New developments in the spectral asymptotics
of quantum gravity}

\author{Giampiero Esposito\dag, Guglielmo Fucci\ddag, 
Alexander Yu Kamenshchik\S, Klaus Kirsten$\|$}

\address{\dag INFN, Sezione di Napoli, and Dipartimento di 
Scienze Fisiche, Complesso Universitario di Monte
S. Angelo, Via Cintia, Edificio N', 80126 Napoli, Italy}

\address{\ddag Department of Physics, New Mexico Institute of
Mining and Technology, Leroy Place 801, Socorro, NM 87801, USA}

\address{\S Dipartimento di Fisica and INFN, Sezione di Bologna,
Via Irnerio 46, 40126 Bologna, Italy; L D Landau Institute for Theoretical
Physics, Kosygin str. 2, 119334 Moscow, Russia}

\address{$\|$ Department of Mathematics, Baylor University, Waco,
TX 76798, USA}

\begin{abstract}  
A vanishing one-loop wave function of the
Universe in the limit of small three-geometry is found, on imposing
diffeomorphism-invariant boundary conditions on the Euclidean 4-ball
in the de Donder gauge. This result suggests a quantum avoidance of the
cosmological singularity driven by full diffeomorphism invariance of
the boundary-value problem for one-loop quantum theory. All of this is
made possible by a peculiar spectral cancellation on the Euclidean
4-ball, here derived and discussed.
\end{abstract}

\section{Introduction}

Since the early eighties there has been a substantial revival of interest
in quantum cosmology, motivated by the hope of obtaining a complete picture
of how the universe could arise and evolve
\cite{Hawk82, Vile84, Vile86, Vile88}. By complete we here mean a
theoretical description where, by virtue of the guiding principles of
physics and mathematics, both the differential equations of the theory 
and the associated boundary (and initial) conditions are fully specified. 
Even though modern theoretical cosmology deals with yet other deep issues
such as dark matter, dark energy \cite{Kame01, Capo05} 
and cosmic strings \cite{Vile05}, the effort of
formulating the appropriate boundary conditions for the quantum state
of the universe \cite{Hawk84}, 
or at least for its (one-loop) semiclassical approximation, plays
again a key role, since the universe might have had a semiclassical origin
\cite{Hawk05}, and the various orders in $\hbar$ 
in the loop expansion describe
the departure from the underlying classical dynamics.

The physical motivations of our research result 
therefore from the following active areas of research:
\vskip 0.3cm
\noindent
(i) Functional integrals and space-time approach to quantum field theory
\cite{DeWi03}.
\vskip 0.3cm
\noindent
(ii) Attempt to derive the whole set of physical laws from invariance
principles \cite{Espo00}.
\vskip 0.3cm
\noindent
(iii) How to derive the early universe evolution from quantum physics;
how to make sense of a wave function of the universe and of 
Hartle--Hawking quantum cosmology \cite{Hart83, Hawk84}.
\vskip 0.3cm
\noindent
(iv) Spectral theory and its physical applications, including functional
determinants in one-loop quantum theory and hence the first corrections 
to classical dynamics \cite{Espo94}.

The boundary conditions that we study are part of a unified scheme for
Maxwell, Yang--Mills and Quantized General Relativity at one loop, 
i.e. \cite{Avra99}
\begin{equation}
\Bigr[\pi {\mathcal A}\Bigr]_{\mathcal B}=0,
\label{(1)}
\end{equation}
\begin{equation}
\Bigr[\Phi(A)\Bigr]_{\mathcal B}=0,
\label{(2)}
\end{equation}
\begin{equation}
[\varphi]_{\mathcal B}=0.
\label{(3)}
\end{equation}
With our notation, $\pi$ is a projector acting on the gauge field 
$\mathcal A$, $\Phi$ is the gauge-fixing functional, $\varphi$ is
the ghost field (or full set of ghost fields) \cite{Espo97}. Both
equation (1) and (2) are preserved under infinitesimal gauge transformations 
provided that the ghost obeys homogeneous Dirichlet conditions as 
in (3). For gravity, we choose $\Phi$ so as to have an operator $P$
of Laplace type on metric perturbations in the one-loop Euclidean theory.

\section{Eigenvalue conditions for scalar modes}

On the Euclidean 4-ball, we expand metric perturbations $h_{\mu \nu}$
in terms of scalar, transverse vector, transverse-traceless tensor
harmonics on $S^{3}$. For vector, tensor and ghost modes, boundary
conditions reduce to Dirichlet or Robin \cite{Espo05a}. 
For scalar modes, one finds eventually the eigenvalues $E=x^{2}$ from
the roots $x$ of \cite{Espo05a}
\begin{equation}
J_{n}'(x) \pm {n\over x}J_{n}(x)=0,
\label{(4)}
\end{equation}
\begin{equation}
J_{n}'(x)+\left(-{x\over 2}\pm {n\over x}\right)J_{n}(x)=0.
\label{(5)}
\end{equation}
Note that both $x$ and $-x$ solve the same equation. For example, at small
$n$ and large $x$, the roots of Eq. (5) with $+$ sign in front of
${n\over x}$ read as (here $s=0,1,...,\infty$)
\begin{equation}
x(s,n) \sim \beta(s,n) \left[1+{\gamma_{1}\over \beta^{2}(s,n)}
+{\gamma_{2}\over \beta^{4}(s,n)}+{\gamma_{3}\over \beta^{6}(s,n)}
+{\rm O}(\beta^{-8})\right],
\label{(6)}
\end{equation}
where
\begin{equation}
\beta(s,n) \equiv \pi \left(s+{n\over 2}+{3\over 4}\right),
\label{(7)}
\end{equation}
and (having defined $m \equiv 4n^{2}$)
\begin{equation}
\gamma_{1}(m) \equiv -{(m-17)\over 8},
\label{(8)}
\end{equation}
\begin{equation}
\gamma_{2}(m) \equiv -{3455\over 384}+2m^{1/2}+{67\over 192}m
-{7\over 384}m^{2},
\label{(9)}
\end{equation}
\begin{equation}
\gamma_{3}(m)={1117523\over 15360}-{115\over 4}m^{1/2}
-{5907\over 5120}m+{3\over 4}m^{3/2}+{421\over 3072}m^{2}
-{83\over 15360}m^{3},
\label{(10)}
\end{equation}
as we have found in the second item of \cite{Espo05b}.

\section{Four generalized $\zeta$-functions for scalar modes}

From Eqs. (4) and (5) we obtain the following integral representations
of the resulting $\zeta$-functions upon exploiting the Cauchy theorem and 
rotation of contour \cite{Espo05a, Espo05b}:
\begin{equation}
\zeta_{A,B}^{\pm}(s) \equiv {(\sin \pi s)\over \pi}
\sum_{n=3}^{\infty}n^{-(2s-2)}\int_{0}^{\infty}dz \; z^{-2s}
{\partial \over \partial z}\log F_{A,B}^{\pm}(zn),
\label{(11)}
\end{equation}
where (here $\beta_{+} \equiv n, \beta_{-} \equiv n+2$)
\begin{equation}
F_{A}^{\pm}(zn) \equiv z^{-\beta_{\pm}}\Bigr(znI_{n}'(zn)
\pm n I_{n}(zn)\Bigr),
\label{(12)}
\end{equation}
\begin{equation}
F_{B}^{\pm}(zn) \equiv z^{-\beta_{\pm}}\left(znI_{n}'(zn)
+\left({(zn)^{2}\over 2} \pm n \right)I_{n}(zn)\right),
\label{(13)}
\end{equation}
$I_{n}$ being the modified Bessel functions of first kind.
Regularity at the origin is easily proved in the elliptic sectors,
corresponding to $\zeta_{A}^{\pm}(s)$ and $\zeta_{B}^{-}(s)$.

\section{Regularity of $\zeta_{B}^{+}$ at $s=0$}

We now define $\tau \equiv (1+z^{2})^{-1/2}$ and consider the uniform
asymptotic expansion (away from $\tau =1$, with notation as in
\cite{Espo05a, Espo05b})
\begin{equation}
z^{\beta_{+}}
F_{B}^{+}(zn) \sim {{\rm e}^{n\eta(\tau)}\over h(n)\sqrt{\tau}}
{(1-\tau^{2})\over \tau}
\left(1+\sum_{j=1}^{\infty}{r_{j,+}(\tau)\over n^{j}}\right),
\label{(14)}
\end{equation}
the functions $r_{j,+}$ being obtained from the Olver polynomials
for the uniform asymptotic expansion of $I_{n}$ and $I_{n}'$ \cite{Olve54}.
On splitting $\int_{0}^{1}d\tau=\int_{0}^{\mu}d\tau
+\int_{\mu}^{1}d\tau$ with $\mu$ small, we get an asymptotic expansion
of the l.h.s. by writing, {\it in the first interval} on the r.h.s.,  
\begin{equation}
\log \left(1+\sum_{j=1}^{\infty}{r_{j,+}(\tau)\over n^{j}}\right)
\sim \sum_{j=1}^{\infty}{R_{j,+}(\tau)\over n^{j}},
\label{(15)}
\end{equation}
and then computing
\begin{equation}
C_{j}(\tau) \equiv {\partial R_{j,+}\over \partial \tau}
=(1-\tau)^{-j-1}\sum_{a=j-1}^{4j}K_{a}^{(j)}\tau^{a}.
\label{(16)}
\end{equation}
The integral $\int_{\mu}^{1}d\tau$ is instead found to yield a vanishing
contribution in the $\mu \rightarrow 1$ limit 
(second item in \cite{Espo05b}).
Remarkably, by virtue of the spectral identity
\begin{equation}
g(j) \equiv \sum_{a=j}^{4j}{\Gamma(a+1)\over \Gamma(a-j+1)}
K_{a}^{(j)}=0,
\label{(17)}
\end{equation}
which holds $\forall j=1,...,\infty$, we find
\begin{equation}
\lim_{s \to 0}s \zeta_{B}^{+}(s)={1\over 6}\sum_{a=3}^{12}
a(a-1)(a-2)K_{a}^{(3)}=0,
\label{(18)}
\end{equation}
and 
\begin{equation}
\zeta_{B}^{+}(0)={5\over 4}+{1079\over 240}-{1\over 2}
\sum_{a=2}^{12}\omega(a)K_{a}^{(3)} 
+ \sum_{j=1}^{\infty}f(j)g(j)={296 \over 45},
\label{(19)}
\end{equation}
where
\begin{eqnarray}
\omega(a) & \equiv & {1\over 6}{\Gamma(a+1)\over \Gamma(a-2)}
\biggr[-\log(2)-{(6a^{2}-9a+1)\over 4}{\Gamma(a-2)\over \Gamma(a+1)}
\nonumber \\
&+& 2\psi(a+1)-\psi(a-2)-\psi(4)\biggr],
\label{(20)}
\end{eqnarray}
\begin{equation}
f(j) \equiv {(-1)^{j}\over j!}\Bigr[-1-2^{2-j}+\zeta_{R}(j-2)
(1-\delta_{j,3})+\gamma \delta_{j,3}\Bigr].
\label{(21)}
\end{equation}
The spectral cancellation (17) achieves three goals: (i) Vanishing of
$\log 2$ coefficient in Eq. (19); (ii) Vanishing of 
$\sum_{j=1}^{\infty}f(j)g(j)$ in Eq. (19); (iii) Regularity at the
origin of $\zeta_{B}^{+}$.

To cross-check our analysis, we evaluate $r_{j,+}(\tau)-r_{j,-}(\tau)$
and hence obtain $R_{j,+}(\tau)-R_{j,-}(\tau)$ for all $j$. Only $j=3$
contributes to $\zeta_{B}^{\pm}(0)$, and we find
\begin{eqnarray}
\zeta_{B}^{+}(0)&=& \zeta_{B}^{-}(0)-{1\over 24}\sum_{l=1}^{4}
{\Gamma(l+1)\over \Gamma(l-2)}\left[\psi(l+2)-{1\over (l+1)}\right]
\kappa_{2l+1}^{(3)} \nonumber \\
&=& {206 \over 45}+2={296\over 45},
\label{(22)}
\end{eqnarray}
in agreement with Eq. (19), where $\kappa_{2l+1}^{(3)}$ are the four
coefficients on the right-hand side of
\begin{equation}
{\partial \over \partial \tau}(R_{3,+}-R_{3,-})
=(1-\tau^{2})^{-4}\Bigr(80 \tau^{3}-24 \tau^{5}+32 \tau^{7}
-8 \tau^{9}\Bigr).
\label{(23)}
\end{equation}
Within this framework, the spectral cancellation reads as
\begin{equation}
\sum_{l=1}^{4}{\Gamma(l+1)\over \Gamma(l-2)}
\kappa_{2l+1}^{(3)}=0,
\label{(24)}
\end{equation}
which is a particular case of
\begin{equation}
\sum_{a=a_{\rm min}(j)}^{a=a_{\rm max}(j)}
{\Gamma((a+1)/2)\over \Gamma((a+1)/2 -j)}\kappa_{a}^{(j)}=0.
\label{(25)}
\end{equation}
Interestingly, the full $\zeta(0)$ value for pure gravity (i.e. including
the contribution of tensor, vector, scalar and ghost modes) is then
found to be positive: $\zeta(0)={142\over 45}$ \cite{Espo05b}, which
suggests a quantum avoidance of the cosmological singularity driven
by full diffeomorphism invariance of the boundary-value problem for
one-loop quantum theory \cite{Espo05b}.

\section{Open problems}

Several open problems should be brought to the attention of the reader,
and are as follows.
\vskip 0.3cm
\noindent
(i) We have encountered a boundary-value problem where the generalized
$\zeta$-function remains well defined, even though the Mellin transform
relating $\zeta$-function to heat kernel does not exist
(see further comments below), since strong 
ellipticity is violated \cite{Avra99} (see also
\cite{Dowk97}). Are the spectral
cancellations (17) and (25) a peculiar property of the Euclidean 4-ball,
or can they be extended to more general Riemannian manifolds with
non-empty boundary?
\vskip 0.3cm
\noindent
(ii) What is the deeper underlying reason for finding
$\zeta_{B}^{+}(0)-\zeta_{B}^{-}(0)=2$? Is it possible to foresee
a geometrical or topological or group-theoretical origin of
this result?
\vskip 0.3cm
\noindent
(iii) Is it correct to say that our positive $\zeta(0)$ value for pure
gravity engenders a quantum avoidance of the cosmological singularity at
one-loop level? \cite{Espo05b, Hawk05} Does the result remain true in
higher-loop calculations or on using other regularization techniques
for the one-loop correction?
\vskip 0.3cm
\noindent
(iv) The whole scheme might be relevant for AdS/CFT in light of a 
profound link between AdS/CFT and the Hartle--Hawking wave function of
the universe \cite{Horo04}.
\vskip 0.3cm
\noindent
(v) What happens if one considers instead non-local boundary data,
e.g. those giving rise to surface states for the Laplacian? 
\cite{Espo00, Schr89, Espo99} 
\vskip 0.3cm
As far as item (i) is concerned, we should add what follows.
The integral representation (11) of the generalized $\zeta$-function
is legitimate because, for any fixed $n$, there is a countable infinity
of roots $x_{j}$ and $-x_{j}$ of Eqs. (4) and (5), and they grow
approximately linearly with the integer $j$ counting such roots. The
functions $F_{A}^{\pm}$ and $F_{B}^{\pm}$ admit therefore a 
canonical-product representation \cite{Ahlf66} 
which ensures that the integral
representation (11) reproduces the standard definition of generalized
$\zeta$-function \cite{Espo05a}. Furthermore, even though the Mellin
transform relating $\zeta$-function to integrated heat kernel cannot 
be exploited when strong ellipticity is not fulfilled, it remains possible
to define a generalized $\zeta$-function. For this purpose, a weaker
assumption provides a sufficient condition, i.e. the existence of a sector
in the complex plane free of eigenvalues of the leading symbol of the
differential operator under consideration \cite{Espo05a, Espo05b}. To make
sure we have not overlooked some properties of the spectrum, we have been
looking for negative eigenvalues or zero-modes, but finding none.
Indeed, negative eigenvalues $E$ would imply purely imaginary roots
$x=iy$ of Eq. (5), but such roots do not exist, as one can check both
numerically and analytically; zero-modes would be non-trivial eigenfunctions 
belonging to zero-eigenvalues, but all modes (tensor, vector, scalar 
and ghost modes) are combinations of regular Bessel functions \cite{Espo05a}
(since we require regularity at the origin of the left-hand side
of Eqs. (1)--(3))
for which this is impossible. As far as we can see, we still find sources
of singularities at the origin in the generalized $\zeta$-function 
resulting from lack of strong ellipticity, but the particular symmetries
of the Euclidean 4-ball background reduce them to the four terms in
Eq. (24), which add up to zero despite two of them are non-vanishing. 

We have proposed to interpret the result $\zeta(0)={142\over 45}$ for
pure gravity as an indication that full diffeomorphism invariance of
the boundary-value problem engenders a quantum avoidance of the 
cosmological singularity. Indeed, on the one hand, the work by
Schleich \cite{Schl85} had found that, on restricting the functional
integral to transverse-traceless perturbations, the one-loop 
semiclassical approximation to the wave function of the universe diverges 
at small volumes, at least for the boundary geometry of a three-sphere.
The divergence of the wave functional does not imply, by itself, that
the probability density of the wave functional diverges at small volumes,
since the probability density $p[h]$ on the space of wave functionals
$\psi[h]$ is given by $p[h]=m[h]|\psi|^{2}[h]$, where $m[h]$ is the 
measure on this space, the scaling of which is not known in general. 
On the other hand, in our manifestly covariant evaluation of the one-loop
functional integral for the wave function of the universe, it seems 
incorrect to assume that the measure $m[h]$ scales in such a way
as to cancel exactly the contribution of the squared modulus of $\psi$,
which is proportional to the three-sphere radius raised to the power
$2\zeta(0)$. Thus, we find that our one-loop wave function of the 
universe vanishes at small volume. The normalizability condition of
the wave function in the limit of small three-geometry, which is
weaker than requiring it should vanish in this limit, was instead 
formulated and studied in \cite{Barv90}.

The years to come will hopefully tell us whether our calculations may
be viewed as a first step towards finding under which conditions a
quantum theory of gravity is singularity free in cosmology \cite{Kief05}.
For this purpose, it might also be interesting to study 
diffeomorphism-invariant boundary conditions for $f(R)$ theories of
gravity, recently studied at one-loop level on manifolds without boundary
\cite{Cogn05}.

\section*{Acknowledgments}

We are grateful to Gerald Dunne, Dmitri Fursaev, Mariel Santangelo
and Antonello Scardicchio for stimulating questions, and to
Emilio Elizalde and Sergei Odintsov 
for having organized such a beautiful Conference.
The work of G Esposito has been partially supported by PRIN {\it SINTESI}.
K Kirsten is grateful to the Baylor University Research Committee, to the
Max-Planck-Institute for Mathematics in the Sciences (Leipzig, Germany) and
to the INFN for financial support. The work of A Yu Kamenshchik was partially
supported by the Russian Foundation for Basic Research under the Grant No
02-02-16817 and by the Scientific School Grant No 2338.2003.2


\section*{References}

\begin{thebibliography}{99}

\bibitem{Hawk82}
Hawking S W 1982 {\it Pont. Acad. Sci. Scr. Varia} {\bf 48} 563

\bibitem{Vile84}
Vilenkin A 1984 {\it Phys. Rev.} {\bf D30} 509

\bibitem{Vile86}
Vilenkin A 1986 {\it Phys. Rev.} {\bf D33} 3560

\bibitem{Vile88}
Vilenkin A 1988 {\it Phys. Rev.} {\bf D37} 888

\bibitem{Kame01}
Kamenshchik A Yu, Moschella U and Pasquier V 2001 {\it Phys. Lett.}
{\bf B511} 265

\bibitem{Capo05}
Capozziello S, Nojiri S and Odintsov S D 2005 Unified phantom cosmology:
inflation, dark energy and dark matter under the same standard 
(hep-th/0507182)

\bibitem{Vile05}
Vilenkin A 2005 Cosmic strings: progress and problems (hep-th/0508135)

\bibitem{Hawk84}
Hawking S W 1984 {\it Nucl. Phys.} {\bf B239} 257

\bibitem{Hawk05}
Hawking S W 2005 {\it Phys. Scripta} {\bf T117} 49

\bibitem{DeWi03}
DeWitt B S 2003 {\it The Global Approach to Quantum Field Theory}
(Clarendon Press, Oxford)

\bibitem{Espo00}
Esposito G 2000 {\it Int. J. Mod. Phys.} {\bf A15} 4539

\bibitem{Hart83}
Hartle J B and Hawking S W 1983 {\it Phys. Rev.} {\bf D28} 2960

\bibitem{Espo94}
Esposito G 1994 {\it Quantum Gravity, Quantum Cosmology and Lorentzian
Geometries}, Lecture Notes in Physics, New Series m, Vol. {\bf m12}
(Springer, Berlin)

\bibitem{Avra99}
Avramidi I G and Esposito G 1999 {\it Commun. Math. Phys.} 
{\bf 200} 495

\bibitem{Espo97}
Esposito G, Kamenshchik A Yu and Pollifrone G 1997 {\it Euclidean
Quantum Gravity on Manifolds with Boundary}, Vol. {\bf 85} of
{\it Fundamental Theories of Physics} (Kluwer, Dordrecht)

\bibitem{Espo05a}
Esposito G, Fucci G, Kamenshchik A Yu and Kirsten K 2005 
{\it Class. Quantum Grav.} {\bf 22} 957

\bibitem{Espo05b}
Esposito G, Fucci G, Kamenshchik A Yu and Kirsten K 2005
(hep-th/0506223; {\it JHEP} 0509:063)

\bibitem{Olve54}
Olver F W J 1954 {\it Phil. Trans. R. Soc. Lond.} 
{\bf A 247} 328

\bibitem{Dowk97}
Dowker J S and Kirsten K 1997 {\it Class. Quantum Grav.}
{\bf 14} L169

\bibitem{Horo04}
Horowitz G T and Maldacena J 2004 {\it JHEP} {\bf 0402} 008

\bibitem{Schr89}
Schr\"{o}der M 1989 {\it Rep. Math. Phys.} {\bf 27} 259

\bibitem{Espo99}
Esposito G 1999 {\it Class. Quantum Grav.} {\bf 16} 1113

\bibitem{Ahlf66}
Ahlfors L V 1966 {\it Complex Analysis} (New York: McGraw--Hill)

\bibitem{Schl85}
Schleich K 1985 {\it Phys. Rev.} {\bf D32} 1889

\bibitem{Barv90}
Barvinsky A O and Kamenshchik A Yu 1990 {\it Class. Quantum Grav.}
{\bf 7} L181

\bibitem{Kief05}
Kiefer C 2005 Quantum gravity: general introduction and recent
developments (gr-qc/0508120)

\bibitem{Cogn05}
Cognola G, Elizalde E, Nojiri S, Odintsov S D and Zerbini S 2005
{\it JCAP} 0502:010 

\end{thebibliography}
\end{document}